\documentclass{emulateapj}

\newcommand{\etal}{et~al.\ }

\newcommand{\ie}{i.e.,\ }
\newcommand{\eg}{e.g.,\ }

\begin{document}

\title{Fast and Efficient Template Fitting of Deterministic Anisotropic Cosmological Models Applied to {\em WMAP} Data}
\author{T. R. Jaffe\altaffilmark{1}, A. J. Banday\altaffilmark{1},
  H. K. Eriksen\altaffilmark{2}, K. M. G\'orski\altaffilmark{3},
  F. K. Hansen\altaffilmark{4} }

\altaffiltext{1}{Max-Planck-Institut f\"ur Astrophysik,
Karl-Schwarzschild-Str.\ 1, Postfach 1317, D-85741 Garching bei
M\"unchen, Germany; tjaffe@MPA-Garching.MPG.DE, 
banday@MPA-Garching.MPG.DE.}

\altaffiltext{2}{Institute of Theoretical Astrophysics, University of
Oslo, P.O.\ Box 1029 Blindern, N-0315 Oslo, Norway; Centre of
Mathematics for Applications, University of Oslo, P.O.\ Box 1053
Blindern, N-0316 Oslo; Jet Propulsion Laboratory, M/S 169/327, 4800
Oak Grove Drive, Pasadena CA 91109; California Institute of
Technology, Pasadena, CA 91125; h.k.k.eriksen@astro.uio.no} 

\altaffiltext{3}{JPL, M/S 169/327, 4800 Oak Grove Drive, Pasadena CA
  91109; California Institute of Technology, Pasadena, CA 91125;
  Warsaw University Observatory, Aleje Ujazdowskie 4, 00-478 Warszawa,
  Poland; Krzysztof.M.Gorski@jpl.nasa.gov}

\altaffiltext{4}{Institute of Theoretical Astrophysics, University of
Oslo, P.O.\ Box 1029 Blindern, N-0315 Oslo, Norway; 
f.k.hansen@astro.uio.no.}

\begin{abstract}
  We explore methods of fitting templates to cosmic microwave background (CMB) data, and in
  particular demonstrate the application of the total convolver
  algorithm as a fast method of performing a search over all possible
  locations and orientations of the template relative to the sky.
  This analysis includes investigation of issues such as chance
  alignments and foreground residuals.  We apply these methods to
  compare Bianchi models of type VII$_{h}$ to \emph{WMAP} first year
  data and confirm the basic result of our 2005 paper.
\end{abstract}



\section{Introduction}

The widely accepted model in cosmology, the so-called concordance
model, posits an isotropic and homogeneous universe with small
anisotropies generated by primordial fluctuations in the inflationary
field.  These anisotropies are present in the cosmic microwave
background (CMB), which should then be statistically isotropic and
Gaussian.  Many CMB studies therefore examine the CMB from a
statistical point of view with the intention of testing for violations
of these properties.  Alternative cosmological models have not,
however, been completely ruled out, and there are several anomalies in
the \emph{Wilkinson Microwave Anisotropy Probe} (\emph{WMAP}) data
that indicate that such models merit further investigation by
alternate means.

We investigate methods for testing any deterministic anisotropic
cosmological model.  The predicted anisotropy template can be compared
to the data using fitting techniques in both pixel and harmonic space
to search for correlations.  We present a description of these methods
and apply a fast and efficient algorithm for searching the full sky
for the best orientation of a template relative to the data.  We test
these methods with both full- and incomplete-sky data sets, and use
simulations to characterize the significance of the results.

Motivated by the morphology of several detected violations of
Gaussianity and/or isotropy in the \emph{WMAP} data \citep{de
Oliveira-Costa:2004, eriksen:2004a, hansen:2004a, vielva:2004}, we
test our methods using Bianchi type VII$_{h}$ models and the
\emph{WMAP} first-year data.  A preliminary analysis was published in
\citet{jaffe:2005}, in which we reported on a surprisingly significant
detection of a Bianchi model at the $99.7\%$ significance level
compared to simulations.  Here we present an improved search of the
model space, confirm the basic result, and discuss in detail issues
such as foreground contamination and chance alignments.

\section[]{Methods}\label{method}

\subsection{Template Fitting}\label{fitting}

Given any anisotropy pattern that contributes to the data as an
additional component of the observed microwave sky (whether
topological in origin, as in the case of Bianchi models, or
foreground), we perform a fit of the template to the \emph{WMAP} data
as has been done in the past by, \eg \citet{gorski:1996} and
\citet{banday:1996} for foreground analysis.  The best-fit amplitude
$\alpha$ for a template vector ${\mathbf t}$ compared to a data vector
${\mathbf d}$ can be measured by minimizing
\begin{equation}
\chi^2 = ({\mathbf d}-\alpha {\mathbf t})^T  {\mathbf
  M}^{-1}_{\textrm{SN}} ({\mathbf d}-\alpha {\mathbf t}) = {\bf
  \tilde{d}}^T  {\mathbf M}^{-1}_{\textrm{SN}} {\bf
  \tilde{d}},
\label{eq:chi2}
\end{equation}
where ${\mathbf M}_{\textrm{SN}}$ is the covariance matrix including both signal
and noise for the template-corrected data vector ${\bf \tilde{d}}
\equiv {\mathbf d} - \alpha {\mathbf t}$.  Solving for $\alpha$ then becomes
\begin{equation}
\alpha = \frac{ {\mathbf t}^T{\mathbf M}^{-1}_{\textrm{SN}} {\mathbf d} }{ {\mathbf t}^T{\mathbf M}^{-1}_{\textrm{SN}} {\mathbf t} }.
\label{eq:cc_basic}
\end{equation}

To compare multiple template components to a given data sets, \eg
different foregrounds, the problem becomes a matrix equation.
In the case in which we have $N$ different foreground components, we
define
\begin{equation}
{\bf \tilde{d}} = {\mathbf d} - \sum_{k=1}^{N} \alpha_k {\mathbf t}_k 
\end{equation}
and 
\begin{equation}
{\mathbf M}_{\textrm{SN}} = \bigl< {\bf \tilde{d}}  {\bf \tilde{d}}^T \bigr> = {\mathbf M}_{\textrm{S}} + \mathbf{M}_{\textrm{N}}.
\end{equation}
In this case, minimizing  ${\bf \tilde{d}}^T  {\mathbf
M}^{-1}_{\textrm{SN}}  {\bf\tilde{d}}$ leads to the following set
of equations,
\begin{equation}
  \sum_{j=1}^N \mathbf{t}^T_k  \mathbf{M}^{-1}_{\textrm{SN}}  \mathbf{t}_j \alpha_j = \mathbf{t}^T_k  \mathbf{M}^{-1}_{\textrm{SN}}  \mathbf{d}. 
\end{equation}
This is the simple system of linear equations ${\mathbf Ax}={\mathbf b}$,
where 
\begin{eqnarray*}
A_{kj}=\mathbf{t}^T_k  \mathbf{M}_{\textrm{SN}}^{-1}  \mathbf{t}_j, 
\\ 
b_k = \mathbf{t}^T_k  \mathbf{M}_{\textrm{SN}}^{-1}  \mathbf{d} 
\\  
x_k =\alpha_k.
\end{eqnarray*}
When only one template is present, this reduces to equation
(\ref{eq:cc_basic}) above.

The errors $\delta\alpha^k_\nu$ are the square root of the diagonal of
${\mathbf A}^{-1}$.  The matrix ${\mathbf A}$ gives information about
the cross-correlation between the templates themselves.  

Note that the above is equally valid in pixel space or harmonic space.
In the former, it is very easy to account for incomplete sky coverage
or to remove, for example the Galactic plane, by simply including in the data
vectors only the relevant pixels, and likewise by including only the
corresponding rows and columns of the covariance matrix.  The noise in
pixel space is usually well represented by a diagonal matrix
representing uncorrelated pixel noise.  But the signal covariance
matrix in pixel space is large and not sparse, which makes harmonic
space more convenient when this is possible.  In harmonic space and under the
assumption of Gaussianity, the signal covariance is diagonal, and with
the approximation of uncorrelated noise that is uniform over the sky,
the noise covariance can be made to be so as well.  The difficulty in
harmonic space is the sky coverage.  As discussed by
\citet{mortlock:2002}, the coupling matrix to cut out the Galactic
plane using a cut the size of $|b|>20\degr$ becomes numerically
singular for resolutions of $\ell_{\textrm{max}}>50$.  Cuts such as
the conservative Kp0 mask, defined by the \emph{WMAP} team, remove more
of the sky and, due to their structure, the coupling matrix is more
difficult to compute.

We define a method that applies harmonic space fitting to the
full-sky cases using highly processed maps discussed in \S\ \ref{maps}.  
This allows us to increase the computational efficiency using the
algorithm described in \S\ \ref{totalc}.  
Here, we use a uniform mean noise approximation that has a diagonal
covariance matrix.  We use pixel-space fitting for each band
separately in the cut-sky analysis in which the Galactic plane region is
masked out.  (At the \emph{WMAP} signal-to-noise ratio level, little would
be gained by simultaneously fitting the different bands, and the
memory and CPU requirements to invert the matrix would become
onerous.)  Again, we use a diagonal noise approximation that this time
takes into account the observation pattern but not the effects of
smoothing.  Comparisons of fits with fully correct noise to those
using these approximations show that the results do not vary
significantly (at most a few percent, or a small fraction of the error
bar).
All codes have been cross-checked with identical inputs to confirm
identical outputs.

Note that the cosmic monopole and dipole are not known, and although a
best-fit dipole is subtracted from the data in the map making process,
small residual monopole and dipole terms remain in the data.  For this
reason, we cannot include this component in the fit.  In harmonic
space, any monopole and dipole terms can simply be excluded or ignored
by setting, \eg $C_1=C_2=10^8\mu \textrm{K}^2$.  In pixel space, we
fit the monopole and dipole simultaneously as independent components.
See
\S
\ref{fg_bias} for discussion.

The above method determines the best-fit amplitude for a given
template at a fixed orientation relative to the sky.  For foreground
fitting, that is generally all that is required, but in the search for
an anisotropic cosmological component, there are two additional
complexities.  First, we have no {\em a priori} guess for where the
symmetry axis may be pointing and must thus search the entire sky.
Section
\ref{totalc} describes the algorithm we use to do this quickly and
efficiently.  Second, we may have an infinite number of possible
templates (e.g., parameterized as described in \S\ 
\ref{models}) among which we want to find the ``best'', so in
addition to determining the best-fit amplitude for each template, we
need a way to compare how well different templates fit the data and to
select the most interesting.
Section \ref{best} discusses how we address this.

\subsection{Total Convolver}\label{totalc}

The search for the best orientation of a template compared to the data
requires that we evaluate the statistic $\alpha$ described in the
previous section at every possible relative orientation of the
template and data.  Working in harmonic space allows us to use an
algorithm based on Fourier transforms to speed up this search
significantly.

In the case of full-sky analysis, the location or orientation of the
template does not affect the error, \ie  $\delta\alpha$ is invariant.
Then the maximum of $\alpha$ is found at the maximum of the numerator
in equation
(\ref{eq:cc_basic}) above, ${\mathbf t}^T {\mathbf
M}^{-1}_{\textrm{SN}} {\mathbf d}$.  Neglecting for the moment the
covariance matrix, the quantity to be maximized is simply the
convolution of the data with the template.  We seek the maximum over
all possible locations and orientations, and this can be found
efficiently using the total convolver algorithm described in
\citet{wandelt:2001}, which was originally developed for map making
using instrument beams.  

This algorithm decomposes the Euler angles into what amounts to a scan
pattern and then takes advantage of the form the convolution takes in
harmonic space to simplify the calculation.  The rotation operator
$D(\Phi_2,\Theta,\Phi_1)$ can be factored into
$D(\phi_{\textrm{E}},\theta_{\textrm{E}},0)D(\phi,\theta,\omega)$,
where a pre-defined scan pattern determines $\theta_{\textrm{E}}$ and
$\theta$, which in the case of full-sky coverage are both $\pi/2$, so
that the set of angles $(\phi_{\textrm{E}},\phi,\omega)$ covers the
full sky at all possible orientations (see \citet{wandelt:2001}
Figure 1).  (In {\em only} this context of total convolution on the
full sky, $\phi$ corresponds to the polar angle and
$\phi_{\textrm{E}}$ to the azimuthal angle.  Elsewhere in this paper,
these are represented by the more common $\theta$ and $\phi$.)
Defining $T(\phi_{\textrm{E}},\phi,\omega)\equiv {\mathbf
  t}^T{\mathbf d}$ as the quantity to be maximized, $b_{\ell m}$ as
the spherical harmonic coefficients of the template ${\mathbf t}$, and
$a_{\ell m}$ as that for the data ${\mathbf d}$, the convolution is then
\citep[eqs.~9~\&~8]{wandelt:2001}
\begin{eqnarray}
\label{eq:tc_basic}
T_{mm'm''} = \sum_l a_{\ell m} d^\ell_{mm'}(\theta_{\textrm{E}}) d^\ell_{m'm''}(\theta) b^*_{\ell m''}  \\
T(\phi_{\textrm{E}},\phi,\omega) = \sum_{m,m',m''} T_{mm'm''}e^{im\phi_{\textrm{E}} + im'\phi + im''\omega},
\end{eqnarray}
where $d^\ell_{mm'}(\theta)$ is the real function such that
$D^\ell_{mm'}(\phi_2,\theta,\phi_1)=e^{-im\phi_2}d^\ell_{mm'}(\theta)e^{-im'\phi_1}$.
The problem has then become simply to calculate $T_{mm'm''}$ and
Fourier transform to $T(\phi_{\textrm{E}},\phi,\omega)$ to find
the maximum.
To take into account the signal and noise covariance, we simply use a
``whitened'' data vector, ${\bf M}_{\textrm{SN}}^{-1}{\mathbf d}$.

The total convolver can find the best-fit position with an accuracy
limited only by the resolution of the inputs.  The positional accuracy
is $\pi/\ell_{\textrm{max}}$, which for our analysis is $2\degr .8$.  Note
that this is larger than the size of a pixel at the usual HEALPix
resolution of $N_{\textrm{side}}=\ell_{\textrm{max}}/2$.

It should also be noted that searching the full sky will {\em not}
return an unbiased estimate for the amplitude.  Simulations with a
known input value for a particular template at a known position will,
on average, have slightly higher amplitudes returned by the search.
If the correct template location is simply fit to an ensemble of
simulations with additional CMB and noise, the returned amplitudes
will have a Gaussian distribution with the correct mean and variance,
but the same is not true when one is searching for the best location
and orientation as well.  This is because the search is seeking the
maximum, and the resulting distribution is a form of extreme value
distribution\footnote{See, e.g., {\tt http://mathworld.wolfram.com/
ExtremeValueDistribution.html}\label{foot:evd}}, which introduces a
small positive bias in the results.  For realistic situations with CMB
and noise in addition to the component we are fitting, the total
convolver is likely to find a maximum amplitude a small distance away
from the true position.  How different the amplitudes and positions
are on average depends on the particular case in question, since it is
a function of how dominant the template is compared to the CMB and
noise, and how much the template structure changes over angular distance,
etc.  This is quantified for the particular case in question in \S
\ref{full_sky_accuracy} using simulations.

This method is approximately 2 orders of magnitude faster than
performing the fit in harmonic space over a grid of individual
rotations one at a time.  The disadvantage is the storage requirement
for the matrix $T$, which increases with the third power of the
resolution and becomes over 2GB for a HEALPix resolution of
$N_{\textrm{side}}=128$ or angular resolution of $42\arcmin$.

\subsection{Best-fit Model and Significance}\label{best}

As mentioned above, when it is not one unique template for which we
are testing but rather a set of possibilities, we need not only to
find the best fit of each to the data but also to find the best fit
among the possible models.  
Depending on how the model space is parameterized, there can be an
infinite number of possibilities.  Previous studies seeking upper
limits on shear and vorticity \citep{kogut:1997,bunn:1996} used two
different statistics to determine the ``best''-fit model.

Given a model, Kogut \etal define the best-fit position and amplitude
in terms of $\Gamma=\alpha/\delta\alpha$.  They used
\emph{Cosmic Background Explorer} (\emph{COBE}) data, for which no full-sky analysis was possible.  In
the case of incomplete sky analysis, the amount of template structure
that is masked changes the significance of the fit.  A large amplitude
in which most of the structure is masked by the Galactic plane cut is not
as interesting as a lower amplitude fit in which the structure is
included.  By finding the maximum not of $\alpha$ but of $\Gamma$,
they attempt to find the most significant fit rather than simply the
maximum amplitude.

Bunn \etal (1996) use a different statistic to accomplish the same effective
selection.  They define $\eta_1 \equiv (\chi^2_0 -
\chi^2_1)/\chi^2_0$, where $\chi^2_1$ is as in equation \ref{eq:chi2},
and $\chi^2_0$ the corresponding statistic for the data by itself,
uncorrected for any anisotropic component.  The difference is then an
indication of how much better the data fit the (statistically
isotropic CMB) theory after correction for the anisotropic model.

Finding the maximum of $\Gamma$ is equivalent to finding the maximum
of $\eta_1$ (although Bunn et al. use a different statistical method).
So for a given model, either statistic can be used to find the
best-fit amplitude and position.  But it becomes more complicated to
compare one model to another in order to determine which model fits
the data better.

The problem with the simple approach, used by \citet{kogut:1997} as
well as in our preliminary analysis \citep{jaffe:2005}, of using
$\Gamma$ or $\eta_1$ to find the best model is that the distribution
of these values for chance alignments is not the same for all models.
Although they are generally quite similar, differences in the tails of
the distributions mean that a given value of $\Gamma$ has a slightly
different significance for different models.  This means that finding
the maximum of $\Gamma$ might have missed other models that are
significant but in which the tail of the distribution does not reach as
high in $\Gamma$.  In other words, the significance of the fit found
in our original result is not incorrect, but it is possible that such
an analysis fails to detect another significant model.

For this more complete analysis, we analyze a set of LILC simulations
\citep{eriksen:2004b,eriksen:2005}, using the above formalism to
characterize the distributions of $\alpha$ values for a given model.
In this analysis, for a given model, we compare the $\alpha$ value
(equivalently $\Gamma$, since $\delta\alpha$ does not change for a
given model on the full sky) for the \emph{WMAP} data against the
ensemble of simulations.  We can then quantify the significance of a
given model fit to the data based on the percentage of LILC
simulations in which the model fits with a lower amplitude.  This gives
clearer indication of which are the most interesting models than a
simple $\Gamma$ or $\chi^2$ statistic.
Comparison of the results using $\alpha$ or $\eta_1$ in this way show
there is little difference between the two in terms of how significant
a given fit to the data is against the simulations.  In the following
analysis, we use the numbers for $\alpha$ only.

\subsection{Visualization:  Cross-Correlation Signal Maps}\label{cc_maps}

It is helpful to be able to visualize what parts of the sky are
driving a particular fit.  To do this, we simply note that the
numerator of equation \ref{eq:cc_basic}, ${\mathbf t}^T{\mathbf
M}^{-1}_{\textrm{SN}} {\mathbf d}$, can be rewritten in pixel space
as $\sum_p [{\mathbf L}^{-1} {\mathbf t}]_p [{\mathbf L}^{-1} {\mathbf
d}]_p$, where ${\mathbf L}$ is the ``square root'' of the covariance
matrix ${\mathbf M}_\textrm{SN}$, or its lower triangular decomposition found
from, for example Cholesky decomposition.  A simple visualization is to turn this
into a map, where each pixel contains the product of $[{\mathbf
L}^{-1}_{\textrm{SN}} {\mathbf t}]$ and $[{\mathbf L}^{-1}_{\textrm{SN}}
{\mathbf d}]$ at that pixel.  This map shows exactly what regions on
the sky drive the fit at a given orientation.  This is particularly
important when certain regions of the sky are known to be
contaminated; these plots show whether or how much those regions
affect the fit.  Examples will be shown in \S\ \ref{two_models}.

\section{Data and simulations}

Here we describe the particular class of models we investigate and the
data sets used in the analysis.

\subsection{Bianchi Models}\label{models}

Bianchi type VII$_{h}$ refers to the class of spatially homogeneous
generalizations of Friedmann universes that include small vorticity
(universal rotation) and shear (differential expansion) components.
(Type VII$_0$ includes the flat Friedmann-Robertson-Walker model, and
VII$_{h}$ includes that with negative spatial curvature as special
cases.)
\citet{barrow:1985} solve the geodesic equations to derive the induced
CMB anisotropy by linearizing the anisotropic perturbations about the
Friedmann models.  Their solution does not include any dark energy
component, which is a significant shortcoming considering the
preponderance of evidence that now points to $\Omega_\Lambda \sim
0.7$.  
But we examine them first as a test of our template-fitting 
methods and second because of the intriguing possibility
that they may explain several anomalies in the data.

Following the prescription in \citet{barrow:1985}, we construct a
template for the anisotropy induced by vorticity ($\omega$) and shear
($\sigma$).  Bianchi type VII$_{h}$ models are parameterized by the
current total energy density $\Omega_0$ and a parameter $x$
\citep{collins:1973}, 
\begin{equation}
 x = \sqrt{ \frac{h}{1-\Omega_0} }, 
\end{equation}
where $h$ is related to the canonical structure constants and is that
to which the type VII$_{h}$ refers (see
\citealt{kogut:1997,bunn:1996,barrow:1985}). This parameter can be
understood as the ratio of the scale on which the basis vectors change
orientation to the Hubble radius (present 
values).  The resulting temperature anisotropy pattern is then
described by
\citep[ eq.~4.11]{barrow:1985}
\begin{eqnarray}
  \frac{\Delta T}{T} = \left(\frac{\sigma}{H}\right)_0 \{ & [ B(\theta_R) + A(\theta_R)  ]\sin(\phi_R)  \nonumber \\
&  \pm  [ B(\theta_R) - A(\theta_R) ]\cos(\phi_R) \}, 
\label{eq:bianchi}
\end{eqnarray}
where $A$ and $B$ are also functions of $x$ and $\Omega_0$ and include
integrals over conformal time that trace the geodesic from the
surface of last scattering to observation.  The angles $\theta_R$ and
$\phi_R$ are not the observing angles; those are rather
$\theta_{\textrm{ob}}=\pi-\theta_R$ and $\phi_{\textrm{ob}}=\pi+\phi_R$.
The sign on the $\cos(\phi_R)$ term (or alternatively, the $\phi_R$ to
$\phi_\textrm{obs}$ transformation) determines the handedness. Then
$\sigma$ determines the amplitude of the fluctuation and $x$ the
pitch angle of the spiral.  The vorticity is then
\begin{equation}
 \left(\frac{\omega}{H}\right)_0 = \frac{  \sqrt{2(1+h)(1+9h)} }{
6x^2\Omega_0 } \left(\frac{\sigma}{H}\right)_0.
\label{eq:vorticity}
\end{equation}
Note that the shear and vorticity values in our original paper 
\citep{jaffe:2005} contain an error in amplitude, although the basic conclusions are
not affected.

Equation \ref{eq:bianchi} can be rewritten as
\begin{equation}
\frac{\Delta T}{T} \propto \cos(\phi_R \pm \tilde{\phi}).
\end{equation}

In other words, for a given $\theta_R$, the temperature variation
follows a $\cos(\phi_R)$ dependence.  The phase shift $\tilde{\phi}$ is
ultimately a function of $\theta_R$ and the two physical parameters,
$x$ and $\Omega_0$.  The result is a spiral pattern with approximately
$N=2/\pi x$ twists.  The smaller the $x$, the smaller the
scale at which the basis vectors change their orientations and the
tighter the resulting spiral.  In the case of $\Omega_0<1$ models,
geodesic focusing leads to an asymmetry wherein the spiral structure
appears compressed in one direction along the rotation axis.


\begin{figure}
\plotone{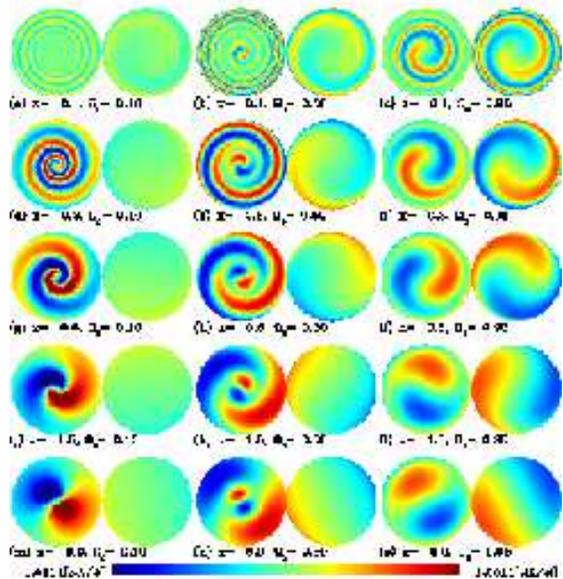}
\figcaption{\small Examples of left-handed Bianchi anisotropy templates in
orthographic projection, all on a common color scale to show the
relative amplitudes.  These must be multiplied by a factor of
$\alpha=(\sigma/H)_0$, \ie the shear (realistically of order
$< 10^{-9}$), in order to give the amplitude of the observed
anisotropy in $\mu \textrm{K}$.  Note that these have been rotated by
$\beta=-90\degr$ to move the center of the structure from the
$-\hat{z}$ pole (as defined in equation
\ref{eq:bianchi}) to the Galactic Center. \label{fig:examples}}
\end{figure}

The template is calculated as $\frac{\Delta T}{(\sigma/H)_0}$, i.e. the
contents of the curly brackets in equation \ref{eq:bianchi} times the
average CMB temperature, so that the shear $(\sigma/H)_0$ is the
amplitude of the template to be found by fitting it against the CMB
anisotropies, $\Delta T$.  Examples are shown in Figure
\ref{fig:examples}, where the template is plotted without the
normalization by the shear.  In generating all of our Bianchi
templates, we have taken the redshift to the surface of last
scattering, or recombination, as $z_{\textrm{rec}}=1100$.  (Changing
to, for example, $z_{\textrm{rec}}=1000$ lowers the amplitude of the
anisotropy by $\sim15\%$, implying a corresponding increase in the
value of the shear $(\sigma/H)_0$ for a given $\Delta T$.)

We make the simple and pragmatic assumption that the anisotropy
induced by the geometry simply adds to the statistically isotropic and
Gaussian component.  

We examine a grid of such models over $0.1\le\Omega_0\le1.0$ in
increments of $0.05$ and over $0.1\le x \le 10.0$ in increments of
$0.05$ in the interval $0.1\le x \le 1.0$ and then
logarithmically sampled up to $x=10.$  A finer grid was also examined
surrounding the best-fit model, $0.52\le x\le 0.68$ and $0.42\le
\Omega_0 \le 0.58$ in increments of $0.02$.  
For the largest values of $x$, the spiral has almost disappeared
(because the scale on which the basis vectors change orientations
becomes larger than the horizon size), and so models of higher $x$
are self-similar.  Smaller values of $x$ start to become physically
unrealistic.
\citet{collins:1973} point out that for $x\sim 0.05$, the
characteristic length scale over which basis vectors change
orientation becomes comparable to the size of large-scale structure,
which means that lower values are ruled out by observations of large
scale homogeneity.  Furthermore, as discussed in \S
\ref{loc_accuracy}, small values of $x$ require higher precision
analysis than is feasible.

\subsection{Data}\label{maps}

For this work, we are interested only in large scale structure.  In
all of the following, unless otherwise noted, we use maps in
HEALPix\footnote{{\tt http://healpix.jpl.nasa.gov/}} format
\citep{healpix} at a resolution of $N_{\textrm{side}}=32$ and smoothed
to an effective beam of FWHM $5\degr.5$, with harmonics up to
$\ell_{\textrm{max}}=64$.

The following full-sky maps are used in this analysis (where all
\emph{WMAP} data products are from the first-year data release):
\begin{itemize}

\item The full-sky \emph{WMAP} Internal Linear Combination (WILC) map released
by the \emph{WMAP} team (see \citealt{bennett:2003b}).  This map is
formed by taking linear combinations of the different bands such that
the foregrounds, each of which has a different spectral dependence
from the CMB, are removed leaving only CMB.  The different weights of
the linear combination are determined solely by the data, via minimum
variance, rather than by any prior assumptions about the foreground
behavior.

\item The Lagrange Internal Linear Combination (LILC) map of
  \citet{eriksen:2004b,eriksen:2005}.  The weights used to form the
  WILC map are slightly sub-optimal with respect to the
  minimum-variance criterion \citep{eriksen:2005}, and this is
  corrected in the LILC map, which uses Lagrange multipliers to
  compute the ILC weights.

\item The foreground-cleaned map of \citet{tegmark:2003}, hereafter
TOH.  This map is also generated by a linear combination of bands,
where in this case, the weights are determined in harmonic space.  

\end{itemize}

All of these maps contain residual foreground emission, some of which
is visible by eye along the Galactic plane and some of which extends
to high latitudes.  It should be noted that none of these maps is
intended for high-precision CMB analysis, but we nevertheless use them
in the following to locate the best-fit Bianchi template by full-sky
convolution.  Simulations show that these fits are affected by two
opposing biases (see \S\ \ref{totalc} and \S\ \ref{loc_accuracy}) that
are larger than the effects of the foreground residuals (see \S
\ref{full_sky_accuracy}),  thus justifying our use of these maps
despite their known disadvantages.  In general, we use the full-sky
maps initially to locate best-fit axis for each Bianchi model (see \S
\ref{full_sky_accuracy}), and then verify the amplitude using
partial-sky algorithms on the following additional data:

\begin{itemize}

\item \emph{WMAP} uncorrected maps for each of the five frequency
bands, co-added from each differencing assembly using noise weighting
(see \citealt{bennett:2003a}) and lso noise-weighted, coadded combinations
of bands Q+V, V+W, Q+V+W, Q-V, V-W, Q-W.


\item Kp0 intensity mask, excluding $23.2\%$ of the pixels in which the
K-band intensity is high and also $0\degr.6$ around known point
sources, downgraded to $N_{\textrm{side}}=32$.

\end{itemize}
Finally, we use observations at other wavelengths as foreground
templates:
\begin{itemize}

\item the \citet{finkbeiner:1999} model for thermal dust emission
(hereafter FDS); 

\item the \citet{schlegel:1998} $100 \mu \textrm{m}$ intensity dust
template (hereafter SFD), which is used an alternative to the FDS
model (see discussion in \S\ \ref{best_result});


\item the \citet{finkbeiner:2003a} H$\alpha$ template, with dust
  correction $f_{\textrm{d}}=0.5$;   

\item the \citet{dickinson:2003} H$\alpha$ template with no dust
correction, which is used as an alternative to the Finkbeiner
template;

\item and the \citet{haslam:1982} 408MHz map of synchrotron emission
processed by \citet{davies:1996}.

\end{itemize}

These foreground components are fit simultaneously to each band over
the incomplete sky using the Kp0 mask, which reduces the effects of
foreground contamination on the fit amplitude (see \S\ \ref{fg_bias}).
Note that although we are simultaneously fitting the foreground
components, these templates are not accurate enough in the Galactic
plane region for full-sky fits to be reliable.

\subsection{Gibbs Samples}\label{gibbs}

In addition to the \emph{WMAP} data products, we also analyze a set of
Gibbs sampled maps that were generated by the method described by
\citet{jewell:2004},\citet{wandelt:2004}, and \citet{eriksen:2004c}.  Effectively, this method samples the space of
CMB signal maps that are consistent with the data, taking into account
both noise characteristics and limited sky coverage. Thus, each single
Gibbs sample represents a full-sky, noiseless CMB signal consistent
with the data assuming Gaussianity, and the distribution of such maps
describes the full CMB signal posterior distribution.

Such sampled maps can thus be analyzed very efficiently using the
total convolver method described above, since neither sky cut nor
non-uniform noise complicate the analysis.  These allow us to avoid
the problem of foreground residuals in the Galactic plane, since this
region of the Gibbs samples contains only CMB signal that is either
consistent with the structure outside the plane, in the case of large
enough scales, or entirely Gaussian random, in the case of smaller
scales.  The ensemble of fit results then reflects how well the
template fits the CMB signal posterior distribution.
In the following, we analyze ensembles of 1000 samples
corresponding to each of the three cosmologically important
\emph{WMAP} Q, V, and W bands.  

\subsection{Assumed Signal Covariance}\label{sig_covar}

Given that we are searching for evidence of anisotropy, the
description of the expected signal covariance is not trivial.  Bianchi
models in particular are not compatible with inflation theory and do
not make any prediction for fluctuations at the surface of last
scattering.  Clearly, a self-consistent theory is required to explain
the observed anisotropies in addition to the Bianchi component, and in
particular, that theory must be consistent with the acoustic peaks now
detected at smaller scales.  No such theory currently exists, but we
note that the Harrison-Zel'dovich power-law spectrum prediction
predates inflation theory.
Because it has been shown to match the data very well on small scales,
we use the inflationary prediction as a starting point.

The signal covariance expected after subtraction of any Bianchi
component is then assumed to be that of Gaussian, isotropic CMB
fluctuations fully characterized by the power spectrum.  We use the
best-fit \emph{WMAP} theoretical power-law spectrum to perform our
fit.  One could then refine the input spectrum based on the result
(\ie do a new parameter estimation using the corrected sky) and
iterate.  In the present analysis, however, we do not aim to improve
the power spectrum estimation.  Template fitting proves to be
insensitive to the assumed power spectrum.  (The fit result changes by
less than $3\%$ when using a flat, $Q=18\mu \textrm{K}$ power spectrum
instead.)
So for the
purposes of this analysis, the best-fit \emph{WMAP} theoretical power-law 
spectrum is sufficient.

\section{Performance, bias, and accuracy}\label{simulations}

In order to interpret the results of the analysis using real data, we
need first to quantify the effects described above.  The model
selection accuracy, the bias due to the maximization over rotations,
any bias due to foreground residuals, and the distribution of chance
alignments are all effects that we can quantify using simulations.

These are generated by the LILC simulation pipeline of
\citet{eriksen:2004b,eriksen:2005}.  The simulations start with a Gaussian CMB
signal generated from an assumed power spectrum and are then smoothed
to the beam width of each \emph{WMAP} differencing assembly.  Pixel noise is
added, uncorrelated and following the instrument properties and
observation pattern described in
\citet{bennett:2003a}.  Finally, the three foreground components above
are added to create simulated raw data for each of the 10 
differencing assemblies.  The LILC algorithm is then used to
reconstruct the corresponding processed, foreground-cleaned sky.
Although these are known to underestimate somewhat the amount of
residual emission along the Galactic plane, they provide a vital
indication of the morphology and approximate amount of such residuals
that may be present in the WILC or LILC maps.

We apply the fitting methods outlined above to the ensemble of LILC
simulations, with and without an additional known anisotropic signal,
to characterize how well the methods perform.  In most of the analysis
below, a set of 1000 simulations were used in the full grid searches
and cut-sky pixel space fitting.  An expanded ensemble of 10,000 LILC
reconstructions was used to refine the significance measures for the
two best-fit models found as described in \S\ \ref{fit_results}.

\subsection{Model Selection Accuracy}\label{model_selection}

First, we add a known Bianchi component (the particular
template and amplitude found in our initial analysis
\citealt{jaffe:2005}) to a set of LILC simulations
and perform the full sky search over all rotations (using the total
convolver) and over the grid of models.  We find that the most
significant model returned is close ($\pm\sim0.1$ in $x$ and
$\Omega_0$) to the correct model in $\sim50\%$ of cases.  Among the
other $\sim50\%$, a qualitatively different model was found to be the
best-fit, but the correct model was still found to be over $99\%$
significant in most cases.  In other words, only in $\sim23\%$ of
realizations was the correct model not detected.

We must then see if we can distinguish the correct model from a false
detection by other means such as incomplete sky fits with simultaneous
foreground template fitting.  These give an idea how much the full-sky
fit is affected by residuals in the Galactic plane.  
%
%
%
Furthermore, models that appear far apart in the model space may in
fact be fitting to the same CMB structure.  We therefore select the
several most significant models to examine in more detail.  Then we
look at what structures are driving the fits and how they behave when
the Galactic plane is excluded and foreground templates simultaneously
fit.  These tools give an additional qualitative way to compare
different model fits.

\subsection{Full-Sky Fitting Accuracy}\label{full_sky_accuracy}

Next, we consider a known Bianchi component added to the input
noiseless, pure CMB realization (as opposed to the LILC
reconstruction) and see how well its position and amplitude are
recovered by the full-sky fit. For 1000 simulations, a Bianchi
component (at the same position and amplitude as our best-fit against
the real data) is added to the input CMB sky and then fit using the
total convolver method described above.  In $\sim80\%$ of
realizations, the returned fit is within $5\degr$ (approximately the
beam width) of the correct location.  
(In the orientation angle, it is less accurate due to the
self-similarity of the spiral structure under such rotations.  The
returned orientation is within $10\degr$ in $52\%$ of the
simulations.)
The amplitudes average $\sim7\%$ {\em higher} than the input value (as
noted in \S
\ref{totalc}), with an rms error of about $80\%$ the calculated error.
Neither of these facts is unexpected, since these values are the
selected maxima, and their distribution is not Gaussian.  The results
are quantitatively the same for the LILC reconstructed skies,
indicating that the foreground residuals do not introduce a
significant additional bias in the case in which a real Bianchi component
is being fitted.  Note that simulations in which the input Bianchi model
has an amplitude a factor of $\sim3$ higher show a much smaller {\it
relative} bias ($\sim1\%$), as one would expect.

\subsection{Cut-Sky Fitting Accuracy}\label{fg_bias}

The cut-sky fits are performed with the Bianchi model at the fixed
location found as the best-fit using the full-sky total convolver
method.  As described in \S\ \ref{totalc}, there is a bias
introduced by the selection of the maximum amplitude position.  This
bias will also be reflected in the cut-sky fits, although masking out
the Galactic plane should remove some of the bias due to residual
foreground emission.

For fits to the raw data outside the Kp0 mask, eight template
components are fit simultaneously to each band,  the three foreground
templates described in \S\ \ref{maps}, a monopole term, the three
spherical harmonics representing the real-valued dipole terms, and the
Bianchi template.

For simulations with no additional Bianchi component, the results show
amplitudes on average $6\%$ lower than those from the LILC fits.  This
is further indication that chance alignments are affected by residuals
in the plane, since the exclusion of that region tends to lower the
fit amplitude.

Simulations with an additional Bianchi component at a known position
and amplitude were run through the same pipeline, \ie first the
full-sky LILC reconstruction was used to find the best-fit location,
then that location used to fit the template to the cut sky in pixel
space.  As described above, the total convolver will return a position
that is very close to the true position but one where the fit amplitude
happens to be highest due to CMB and noise contributions.  These will
also affect the cut-sky fits, which also show a bias of $\sim3\%$.
This is lower than the bias in the full-sky fits, showing that a few
percent of the full-sky bias is due to residuals in the Galactic plane
region.  The relative drop in amplitude between the full- and cut-sky
fits for true detections is on average half the drop in the case of
chance alignment detections.
%
%

\begin{figure}
\plotone{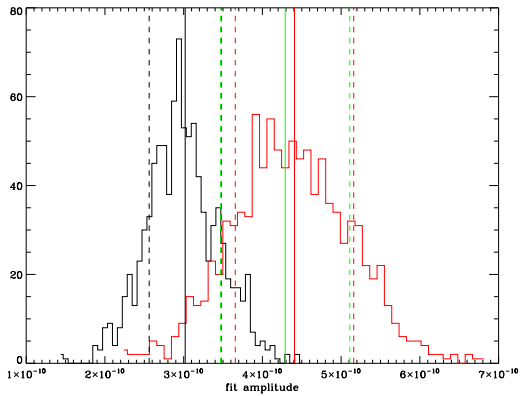}
\figcaption{\small Distributions of fit results for the Bianchi component
for 1000 simulations without (\emph{black lines}) and with (\emph{red lines}) a Bianchi component
added.  Vertical solid lines show the means, and vertical dashed show the
actual errors.  The vertical green line shows the true value and
expected errors.  \label{fig:fit_histos}}
\end{figure}

Figure \ref{fig:fit_histos} shows what these distributions look like
for the fit to 1000 simulations in the V band, both in the case where
a Bianchi component is added (\emph{red histogram}) and where it is
not (\emph{black histogram}).  Also plotted as vertical lines are the
mean and rms errors on the distributions, and the true
value and expected errors plotted in green.  The small bias in the value of the Bianchi
fit is seen in the distance between the vertical red and green lines.

Note that in all these cases, the bias in the fits affects the
absolute amplitude (\ie shear) estimate, but not the significance of
the fit, since the ensemble of simulations used to estimate the
significance is also affected by such a bias.  The expected bias in
the amplitude is also much smaller than the error bar.  Therefore this
does not affect our basic results, namely the particular best-fit
model, its location, its approximate amplitude, and its approximate
significance relative to chance alignments.

\subsection{Chance Alignments}\label{chance_alignments}

For a given sky realization, we find the best model as described in
\S\ \ref{best} and then simply compare the amplitude of that fit
against the ensemble of amplitudes for that model relative to Gaussian
simulations to estimate the significance.  Visual inspection of the
\emph{WMAP} sky maps shows no obvious Bianchi component, so any such
signal must remain at or below the level of the stochastic component.
Chance alignments may therefore either {\em cancel} a Bianchi-induced
signal or give a false positive .  The former effect was quantified in
\S\ \ref{model_selection} at $\sim23\%$, but the latter is more
difficult to quantify.

The family-wise error rate (FWER), the expected number of false
detections when testing $m$ hypotheses, is $\sim m p$ when $p$ is
the probability of one false detection.  If $3\sigma$ is the detection
threshold (implying $p\sim0.003$) and one tests 100 different
hypotheses (or models), the FWER is then $0.3$, meaning one gets a
false detection somewhere in the model space one-third of the time.
Over our grid of Bianchi parameters, the models are not independent
(since models close in $(x,\Omega_0)$ space will resemble each other
closely), so we cannot determine {\em a priori} what the true
frequency of false detections would be, but we can get this from the
ensemble of Gaussian simulations.

We perform the full-sky search using the total convolver over the grid
of Bianchi models and find the best-fit model for each realization.
We find that a false detection due to a chance alignment that has a
significance of $99.7\%$ occurs in $\sim17\%$ of the cases.  A better
comparison might be to use the $\chi^2$ representing the goodness of
the fit.  We then compare the statistic $\eta_1 \equiv (\chi^2_0
-\chi^2_1)/\chi^2_0$ (defined above in \S \ref{best}) , namely the
relative improvement in the $\chi^2$ when the Bianchi model is
subtracted.  We find that by this measure $\sim10\%$ of the best
chance alignments fit their respective realizations as well as our
best-fit model does the \emph{WMAP} data (see \S\ \ref{fit_results}).
Note, however, that these statistics are dependent on the assumed
amount of large-scale power.
The above numbers simply imply that a detection of a Bianchi model
with an amplitude higher than in $99.7\%$ of simulations is more than
4 times as likely to be real as it is to be a chance alignment, in
the absence of all other information.


\section{Application to the first-year \emph{WMAP} data}\label{fit_results}

Armed with the information gained from the analyses of simulations, we
can now examine the fits to the real data.

\subsection{Fits Over Model Space Grid}\label{results_full}

\begin{figure}
\plotone{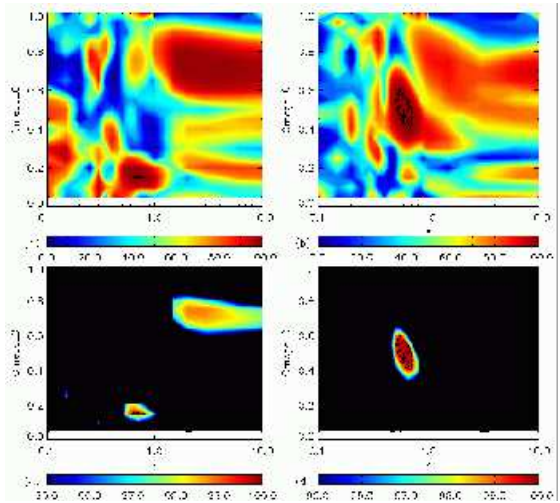}
\figcaption{\small Significance as percentage of LILC simulations whose
best-fit chance alignment amplitude is lower.  Left panels show 
left-handed models, and right panels show right-handed models.  Over plotted contours are at $99.3\%$,
$99.5\%$, $99.7\%$, and $99.9\%$.  Two color scales are used to show the
global structure ([empha{top panels}) as well as that near the peaks (\emph{bottom panels}).\label{fig:contours}}
\end{figure}

Using the total convolver to find the best orientation, we fit the
grid of Bianchi models to each of the WILC, LILC, and TOH full-sky
processed maps.  Figure \ref{fig:contours} shows filled contours over
this grid for the LILC.  (The results for the WILC and TOH look very
similar.)  For each point on the grid corresponding to a model of the
given $(x,\Omega_0)$, the template is fit to the LILC map, and the
color indicates the significance estimate of the resulting amplitude,
\ie the fractional number of Gaussian LILC simulations (out of 1000)
with lower amplitude.  As discussed in \S\ \ref{best}, we use a finer
grid and better method for determining the best-fit model and thereby
select a slightly different model than the analysis in
\citet{jaffe:2005}.  But it is apparent from the right panels of
Figure \ref{fig:contours} that the significance as a function of the
Bianchi parameters $x$ and $\Omega_0$ is flat in the region 
$\pm 0.1$ in both $x$ and $\Omega_0$ about the maximum.  

We find that the most significant fit is found with a right-handed
Bianchi template of $x=0.62$ and $\Omega_0=0.5$ when that template is
rotated to a position and orientation given by Euler angles (following
the total convolver's ``$zyz$'' convention about fixed axes)
$(\Phi_2,\Theta,\Phi_1)=(42\degr,28\degr,-51\degr)$.
As defined in \S\ \ref{models}, the spiral structure of the
unrotated model is centered on the south pole (or $-\hat{z}$ axis), so
this rotation places the center of that structure at Galactic
longitude and latitude of $(l,b)=(222\degr,-62\degr)$ and changes it's
orientation about that location by $\Phi_1=-51\degr$.  This model fits
at an amplitude of $\left(\frac{\sigma}{H}\right)_0=4.29\times
10^{-10}$, which is higher than $99.7\%$ of the 10\,000 simulations.  
This model and the best-fit from previous work \citep{jaffe:2005} at
$x=0.55$ are almost identical.

All models near this best-fit $(x,\Omega_0)$ return the same location
for the center of the spiral within $3\degr$ but vary the orientation
(Euler angle $\Phi_1$) up to $36\degr$. The broad spiral in all of
these models is very self-similar under these rotations, so the change
is driven largely by the precise locations of the paired hot and cold
spots.

Looking at Figure \ref{fig:contours}, one can see that more than one
model appears ``significant'' in the sense of fitting with an
amplitude above $99\%$ of the amplitudes found fitting that
same model to Gaussian simulations.  As discussed above in \S
\ref{model_selection}, this is not surprising, and we must examine each
of these models in more detail.

%
\begin{deluxetable}{lcccc}
\tablecaption{Fitted template amplitudes \label{tab:fits_summary} }

\tablewidth{0pt}
\tablecolumns{4}
\tablehead{  & $(\sigma/H)_0$ & $(\omega/H)_0$ &  $P\left(|\alpha_{\textrm{sim}}| < |\alpha_{\textrm{obs}}|\right)$   \\
Map  & ${\scriptstyle (\times10^{-10})}$ & ${\scriptstyle (\times10^{-10})}$ & \%  }
\startdata
%
%
\cutinhead{Right-handed $(x,\Omega_0)=(0.62,0.5)$} \\
%
WILC & $ 4.33\pm 0.82$ & $ \phm{-}9.58$ & $ 99.8$ \\
LILC & $ 4.29\pm 0.82$ & $ \phm{-}9.49$ & $ 99.7$ \\
TOH & $ 4.03\pm 0.82$ & $ \phm{-}8.92$ & $ 98.6$ \\
%
K\tablenotemark{a} & $ 2.59( 4.13)\pm 0.83$ & $\phm{-} 5.72$ & $ 16.7( 99.1)$ \\
Ka\tablenotemark{a} & $ 3.50( 4.09)\pm 0.83$ & $\phm{-} 7.74$ & $ 86.9( 99.0)$ \\
Q\tablenotemark{a} & $ 3.76( 4.11)\pm 0.83$ & $\phm{-} 8.31$ & $ 95.6( 99.1)$ \\
V\tablenotemark{a} & $ 3.99( 4.19)\pm 0.83$ & $\phm{-} 8.82$ & $ 98.1( 99.5)$ \\
W\tablenotemark{a} & $ 4.08( 4.35)\pm 0.82$ & $\phm{-} 9.03$ & $ 99.1( 99.8)$ \\
QVW\tablenotemark{a} & $ 3.84( 4.15)\pm 0.83$ & $\phm{-} 8.49$ & $ 96.8( 99.2)$ \\
VW\tablenotemark{a} & $ 3.99( 4.22)\pm 0.83$ & $\phm{-} 8.84$ & $ 98.2( 99.6)$ \\
Q-V\tablenotemark{a} & $ 0.06( 0.11)\pm 0.02$ & $\phm{-} 0.13$ & $ 99.0(100.0)$ \\
V-W\tablenotemark{a} & $-0.05(-0.08)\pm 0.02$ & $\phm{-} 0.11$ & $ 93.8( 99.0)$ \\
Q-W\tablenotemark{a} & $ 0.01( 0.04)\pm 0.02$ & $\phm{-} 0.02$ & $ 25.0( 83.2)$ \\

%
Q$^{b}$ & $ 4.09\pm 0.10^c$ & $\phm{-}9.04$ & - \\
V$^{b}$ & $ 4.11\pm 0.10^c$ & $\phm{-}9.09$ & - \\
W$^{b}$ & $ 4.12\pm 0.11^c$ & $\phm{-}9.12$ & - \\

\cutinhead{Left-handed $(x,\Omega_0)=(0.62,0.15)$} \\
%
WILC & $ 2.39\pm 0.47$ & $ \phm{-}22.31$ & $ 97.8$ \\
LILC & $ 2.49\pm 0.47$ & $ \phm{-}23.29$ & $ 99.4$ \\
TOH & $ 2.45\pm 0.47$ & $ \phm{-}22.94$ & $ 99.0$ \\
%
K\tablenotemark{a} & $ 2.33( 3.31)\pm 0.50$ & $\phm{-}21.76$ & $ 96.3( 99.9)$ \\
Ka\tablenotemark{a} & $ 2.24( 2.63)\pm 0.50$ & $\phm{-}20.93$ & $ 94.8( 99.3)$ \\
Q\tablenotemark{a} & $ 2.29( 2.50)\pm 0.50$ & $\phm{-}21.42$ & $ 96.0( 98.9)$ \\
V\tablenotemark{a} & $ 2.33( 2.44)\pm 0.50$ & $\phm{-}21.81$ & $ 96.7( 98.6)$ \\
W\tablenotemark{a} & $ 2.32( 2.46)\pm 0.49$ & $\phm{-}21.69$ & $ 96.3( 98.4)$ \\
QVW\tablenotemark{a} & $ 2.30( 2.48)\pm 0.50$ & $\phm{-}21.46$ & $ 96.1( 98.8)$ \\
VW\tablenotemark{a} & $ 2.34( 2.44)\pm 0.50$ & $\phm{-}21.85$ & $ 96.7( 98.6)$ \\
Q-V\tablenotemark{a} & $ 0.03( 0.02)\pm 0.02$ & $\phm{-} 0.27$ & $ 78.6( 63.1)$ \\
V-W\tablenotemark{a} & $-0.06(-0.10)\pm 0.02$ & $\phm{-}0.55$ & $ 96.4( 99.9)$ \\
Q-W\tablenotemark{a} & $-0.03(-0.08)\pm 0.02$ & $\phm{-}0.29$ & $ 73.1( 99.3)$ \\

%
Q$^{b}$ & $ 2.10\pm 0.11^c$ & $\phm{-}19.67$ & - \\
V$^{b}$ & $ 2.08\pm 0.11^c$ & $\phm{-}19.46$ & - \\
W$^{b}$ & $ 2.09\pm 0.09^c$ & $\phm{-}19.53$ & - \\

\enddata
\tablecomments{ Amplitudes of the best-fit model derived from various
  combinations of data and various methods as described in the
  text. The full sky was used in the analysis of the WILC, LILC, TOH,
  and Gibbs samples, while the Kp0 mask was imposed for the remaining
  maps.}

\tablenotetext{a}{Simultaneous fits with foreground components.  In
parentheses are the values using the SFD dust template instead of the
FDS, and the \cite{dickinson:2003} H$\alpha$ instead of
\citet{finkbeiner:2003a}.}
\tablenotetext{b}{Average over 1000 Gibbs samples.}
\tablenotetext{c}{Errors are rms variation over Gibbs samples.}
%
\end{deluxetable}

\begin{figure}\begin{centering}
%
\epsscale {.70}
\plotone{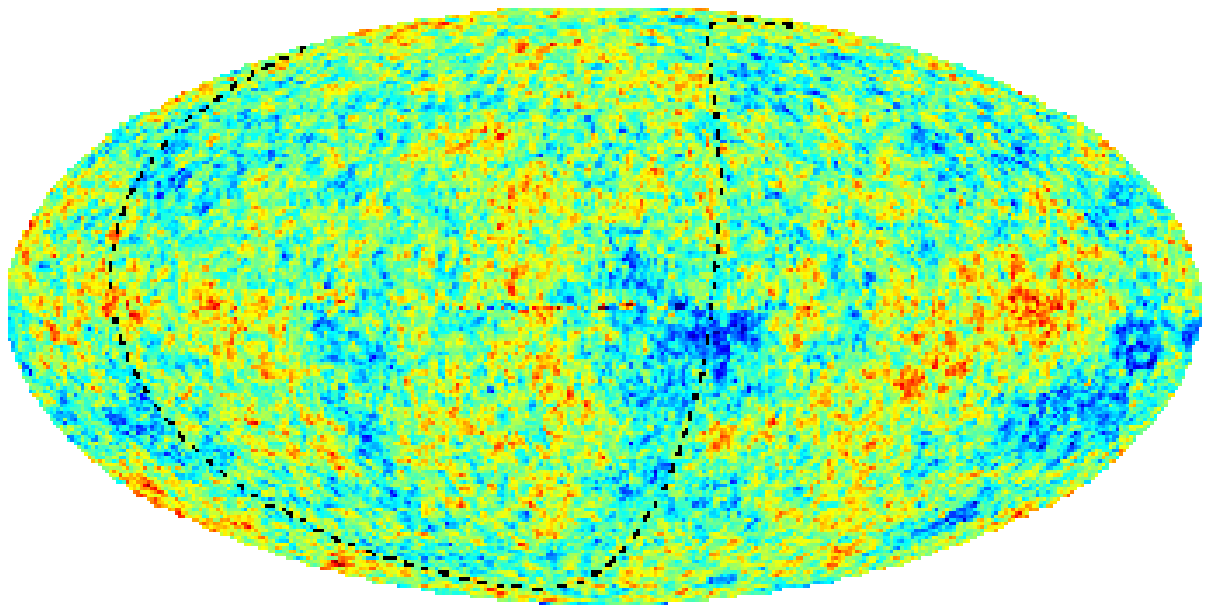}
\plotone{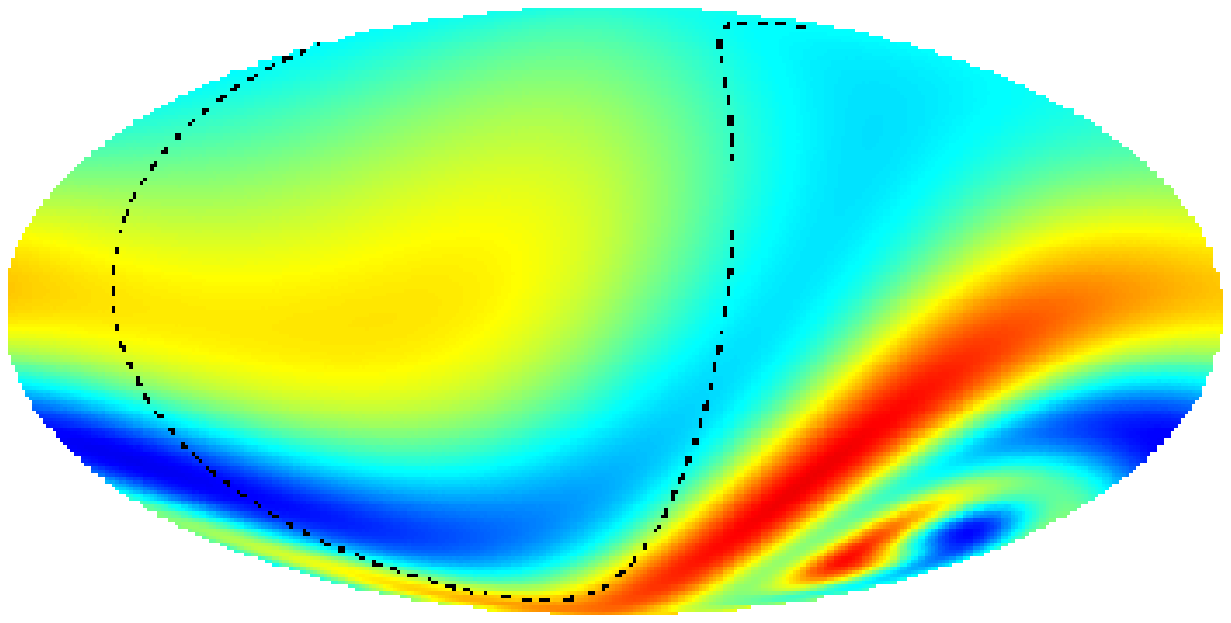}
\plotone{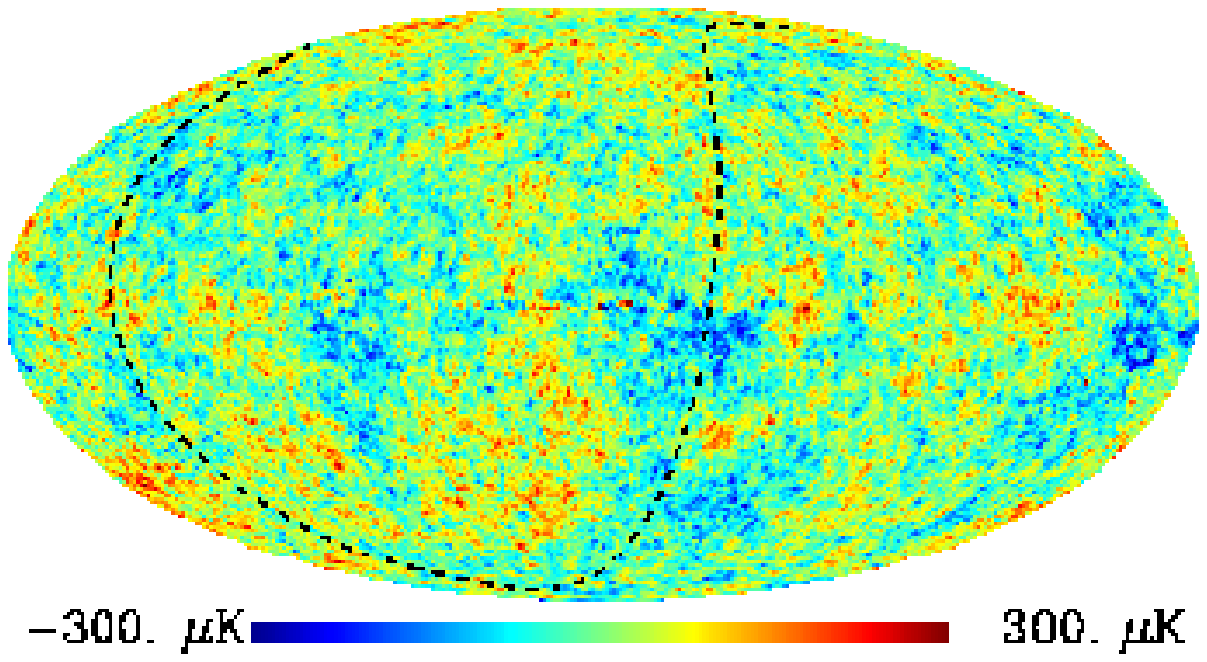}
\figcaption{\small  \emph{Top}:  \emph{WMAP} Internal Linear Combination map. \emph{Middle}:  Best-fit Bianchi VII$_{h}$ template (enhanced by a factor of
  4 to bring out structure). {\it Bottom}:  Difference between WILC
  and best-fit Bianchi template; the ``Bianchi-corrected'' ILC map.
  Over-plotted on each as a dotted line is the equator in the reference
  frame that maximizes the power asymmetry as described in \S\ 
\ref{sec:asymmetry}. \label{fig:ilc_b_bilc}}
\end{centering}
\end{figure}

The full resolution ILC map is shown along with the best-fit Bianchi
model on the same scale and the corrected ILC map in Figure
\ref{fig:ilc_b_bilc}.
A summary of all fit results is shown in Table \ref{tab:fits_summary}.
The expected bias in these results is discussed in \S
\ref{fg_bias}.  The following sections describe the two most
interesting models in more detail.

\subsection{Two Best Fits}\label{two_models}

%
%

\subsubsection{Left-handed Model $(x,\Omega_0)=(0.62,0.15)$}

The most significant left-handed model, at $99.4\%$, is at
$(x,\Omega_0)=(0.62,0.15)$.  This model was not found in our earlier
work \citep{jaffe:2005}, because it is only in a fairly small region
of the model space that this fits with any significance, and our
previous, coarser grid effectively straddled the peak in $\Omega_0$.
The best-fit location for this model puts the center of the structure
at $(l,b)=(320,-20)$, which is closer to the Galactic Center region
than the best-fit right-handed model, raising the question of how
much it
is driven by foreground residuals.

Cut-sky fits give fit amplitudes for this component that are 8\%
lower and significances of  $\sim96\%$ in most cases.  Furthermore,
the Galactic center region tends to draw the template in simulations;
the best-fit location among the simulated LILC maps for this model is
twice as likely to be found in the area around $(0\degr,-20\degr)$ as
should be expected from a uniform distribution.  The only thing that
all of the LILC simulations have in common is foregrounds, so this is
an indication that there is some residual there that is a weak
attractor.  One possibility is the ``free-free haze'' described by
Finkbeiner (2004, see also Patanchon et al.\ 2005), although 
this haze does not match up well with the template structure, the two
show little cross-correlation, and inclusion of Finkbeiner's haze
template in the simultaneous fitting does not alter the fit amplitude
of the Bianchi model.

\begin{figure}\begin{centering}
\epsscale{1.0}
\plotone{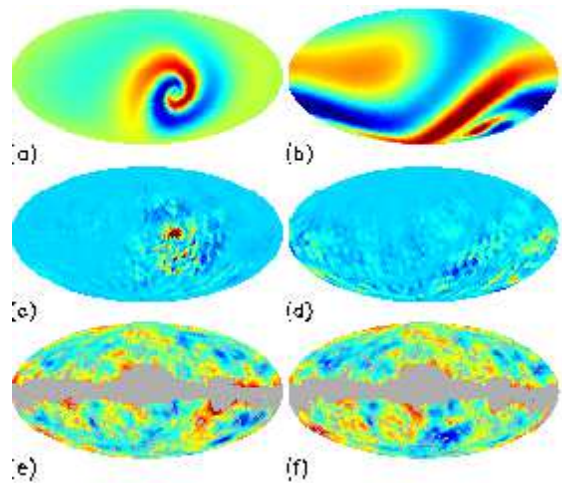}
\figcaption{\small   Two significant models.  The left panels show the
best-fit left handed model with $(x,\Omega_0)=(0.62,0.15)$, while the
right panels show the best-fit overall model, right-handed with
$(x,\Omega_0)=(0.62,0.5)$.  The top panels show the template
amplified by a factor of three to bring out the structure.  The middle
panels show the corresponding cross correlation map (see \S
\ref{cc_maps}) scaled from $-1\%$ to $2\%$.  The bottom panels show the 
``corrected'' \emph{WMAP} Q+V+W map scaled from $-150$ to $150\mu K$.
The grey region is the excluded region of the Kp0 mask.\label{fig:three_fit_maps}}
\end{centering}
\end{figure}

In Figure \ref{fig:three_fit_maps}, it looks like the fit should be
largely driven by the cold region below the Galactic center.  The
cross-correlation maps described in \S\ \ref{cc_maps} do show
correlation there but also indicate that the fit is largely driven by
a very strong signal in the Galactic plane.  Figure
\ref{fig:three_fit_maps} shows these maps for both this model and the
best-fit right handed model.  Where the right handed model shows
relatively uniform correlation over the hemisphere about the best-fit
axis, this model shows a rather concentrated region including a very
strong driver on the Galactic plane.

The Gibbs samples throw further doubt on this model.  Among the 1000
Gibbs samples in each of Q, V, and W bands, this model fits at the
same approximate location as for the LILC map less than half of the
time.  Where the location was the same, the amplitude of the best-fit
is significantly lower for the ensemble of Gibbs-sampled maps, which
drop over $15\%$ in amplitude to a mean of $2.1\times10^{-10}$,
indicating that some of the structure in the data that drives the fits
is not consistent with the posterior CMB distribution as determined by
the Gibbs sampling technique.  Furthermore, this model is almost as
likely to fit near the location of the best-fit right-handed model
instead of near the Galactic center.  This is largely driven by the
cold spot.

In summary, this model is quantitatively less significant than the
best-fit right-handed model based on the cut sky and Gibbs sample fit
values.  Furthermore, the morphology indicates that foreground
residuals drive the full-sky fit.

\subsubsection{Right-handed Model $(x,\Omega_0)=(0.62,0.5)$}\label{best_result}

Figure \ref{fig:contours} shows that the best-fit model is this
right-handed model.

The amplitude of the best-fit Bianchi component varies somewhat across
the different frequencies, in all cases lower than the full-sky
amplitude fit with the LILC.  As discussed in \S
\ref{fg_bias}, this is likely due to small foreground residuals, but does not mean that the detection is a false positive;  the 
same effect occurs in simulations that include a Bianchi component.
The amplitude in the W band, in which the least foreground residuals are
expected, is still higher than $\sim99\%$ of simulations.  The K and
Ka band fits are significantly lower when the FDS dust and Finkbeiner
H$\alpha$ templates are used, but are consistent with the other bands
when the SFD dust and Dickinson H$\alpha$ templates are used instead.
It is known that foreground subtraction is a problem even at high
latitudes in the K and Ka bands, and these residuals are clearly
affecting the low frequency fits.  Looking at the residuals of the two
fits shows that the difference may be driven by a small region around
$(l,b)=(300\degr,-15\degr)$ where the dust templates differ strongly.
The higher frequency fits, however, are more consistent.  The
difference maps, \eg Q-V, should contain no CMB component but only
foreground residuals and noise.  The fact that the Bianchi component
amplitude found from these maps is less than $2\%$ of the co-added map
amplitude is an indication that such residuals are not contributing
significantly to the fit.

The results of fitting the Gibbs-sampled maps show that for this
model, the amplitude is quite stable over the ensemble of Gibbs
samples, with, \eg a mean of $(4.12\pm0.1)\times10^{-10}$ in the W band
compared to $4.08\times10^{-10}$ for the cut-sky fit to the raw data.
Since the Gibbs samples represent the posterior CMB distribution,
taking into account foregrounds and iterating over the power spectrum,
these results are a strong indication that the fit is due primarily to
CMB signal.  

Figure \ref{fig:three_fit_maps} shows the cross-correlation map as
described in \S\ \ref{cc_maps}, which give a visual indication of what
regions drive the fit.  Unlike the left-handed model (\emph{left}), which
shows one concentrated region in the Galactic plane to be driving the
fit, this model correlates over more than half the sky at moderate
levels.  One can see that the cold spot does partly drive the fit, but
no particular region can be said to dominate.  
Fits to the combined QVW and VW data where the cold spot is excluded
(in a $10\degr$ radius around $(l,b)=(209,-57)$) have comparable
amplitudes to fits where the region is included (only $6\%$ lower) .
Further Gibbs samples were also computed while masking this region.
Full sky searches using these samples show that fewer than $20\%$
return positions more than $10\degr$ from the original location, and
amplitudes that are on average $15\%$ lower (which is within the
calculated error bar).  These results confirm that the cold spot does
affect but does not exclusively drive the fit amplitude.

\subsection{Location and Orientation Accuracy}\label{loc_accuracy}

As mentioned above, where a Bianchi component was added to simulations
at a known location, the full-sky search with the total convolver
returned the correct position within $5\degr$ in $\sim80\%$ of
realizations.  The uncertainty in the location is due to the CMB
fluctuations, which are quite comparable to the Bianchi component at
the amplitude detected.  

\begin{figure}
\plotone{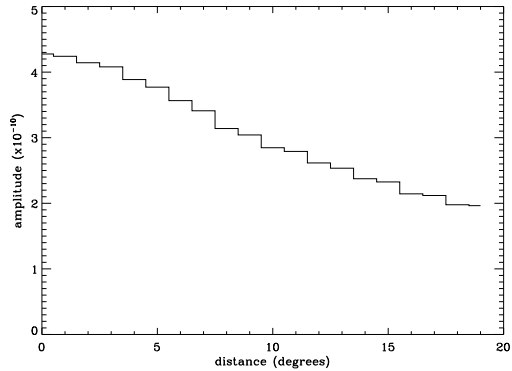}
\figcaption{\small Average fit amplitude as the location
of the template varies from the best-fit position.  For these fits,
the orientation of the template is unchanged.\label{fig:sig_by_dist}}
\end{figure}

\begin{figure}
\plotone{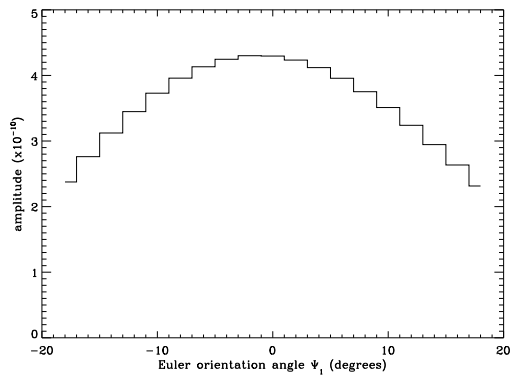}
\figcaption{\small Fit amplitude as the orientation of the
template, \ie the Euler angle $\gamma$, is changed.  For these fits,
the location in longitude and latitude is unchanged.  Note that this
grid finds a preferred orientation $2\degr$ from that found by the
total convolver (due to the slightly different grids used.) \label{fig:sig_by_alpha}}
\end{figure}

To determine how the amplitude changes with the position and
orientation of the template compared to the data, we take the best-fit
Bianchi model and fit it to the LILC on a grid of fixed positions
within $20\degr$ of the best-fit position.  Results are shown in
Figures
\ref{fig:sig_by_dist} and
\ref{fig:sig_by_alpha}.  The orientation is not very sensitive in this
model, whose spiral structure is self similar under rotations about
its symmetry axis; only the precise positions of the hot and cold
spots affect the variation with orientation angle.  The amplitude
drops by $1\%$ when the orientation is $4\degr$ off.  The location of
the symmetry axis is a bit more sensitive, where the amplitude drops
by $3\%$ at $2\degr$.  The fact that the total convolver at this
resolution uses steps of $2\degr.8$ means that its best-fit
amplitude can be several percent off of the actual maximum.  All the
fits to the simulations as well as the data are subject to this same
uncertainty.  If we assume the worst, that the LILC amplitude was
found at its true maximum (\ie the true axis of symmetry happened to
lie exactly on the center of one of the total convolver's bins) and
the simulations are all at $1\degr .4$ away from their true maxima (\ie the
axis exactly between bins) and have true values correspondingly
higher, the comparative significance could then be over-estimated by
only $0.5\%$.  The likely effect is of course much smaller.
 
%

Using the LILC map at higher resolution, $\ell_{\textrm{max}}=128$, gives an
accuracy in the total convolver of $\pi/\ell_{\textrm{max}}=1\degr .4$.  The
position returned is identical, with only the orientation one step of
$1\degr .4$ different.  

The above applies to the best-fit model at $(x,\Omega_0)=(0.62,0.5)$,
but other models have structure at different angular scales.  In
particular, for the region of small $x$ and $\Omega_0$, where a
tightly wound spiral is even more tightly focused in one hemisphere,
the fit amplitudes are far more dependent on the exact position.
Because the total convolver resolution is $\pi/\ell_{\textrm{max}}$, our
analysis is not as sensitive for this region of model space as it
would be for a higher resolution analysis.  In these cases, the
difference of a few degrees can mean a large difference in amplitude.
Simulations show that, although the location returned is the closest bin
to the true location, the amplitude of a model
$(x,\Omega_0)=(0.1,0.1)$ is underestimated by $\sim20\%$ on average
due to the limited resolution.  Increasing the resolution of the
analysis to HEALPix $N_{\textrm{side}}=64$ increases the mean and
brings it closer to the correct value, but it is still underestimated.
(Higher resolution analysis with the total convolver is not feasible
due to the memory and CPU requirements.)
In the region of model space where $x>0.25$ and $\Omega_0>0.25$, this
effect drops to less than a few percent.

A more detailed look at these models at increased resolution
($N_{\textrm{side}}=64$) shows no evidence that the lower resolution
analysis missed a significant detection.  But the limits placed on
shear and rotation are less stringent than they would be were a higher
resolution analysis feasible.

\subsection{DMR Fit}

Our best-fit amplitude is below the upper limit DMR could place on the
shear.  Using this model, a fit to the DMR data gives
$\left(\frac{\sigma}{H}\right)_0=3.38\pm.98\times10^{-10}$, which is
within our best-fit error bar for the \emph{WMAP} data, but which is
not distinguishable from a chance alignment for DMR.
\citet{kogut:1997} report a distribution of $\Gamma$ values for chance
alignments up to $4.5$.  Our fit value and error give $\Gamma=3.4$,
and although this value comes from different methods and assumptions, it
is roughly comparable.

\subsection{Sensitivity to Assumed Power Spectrum}\label{assumptions}

As mentioned in \S\ \ref{sig_covar}, assumptions about the
cosmological parameters go into this analysis from the beginning with
the choice of the signal covariance matrix.  In effect, we are
assuming that the CMB signal consists of an anisotropic
Bianchi-induced component plus a statistically isotropic, Gaussian
random field described completely by its power spectrum, which is 
taken to be the \emph{WMAP} best-fit theoretical power law spectrum.
As we are searching for evidence of a model that affects the power
spectrum at large scales and that is inconsistent with inflation,
this approach obviously lacks consistency.  

\begin{figure}\begin{centering}
\plotone{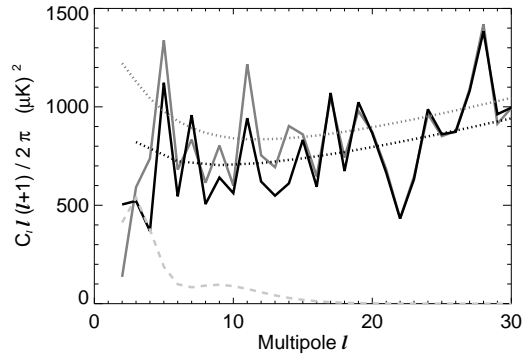}
\figcaption{\small Comparison of power spectra. The gray and black solid lines show the power spectrum estimated from the co-added V+W map before and
after correcting for the Bianchi template, respectively.  The 
dotted gray and black lines shows the theoretical best-fit
power-spectra from the \emph{WMAP}-team analysis and
\citet{hansen:2004b} respectively. The latter is a fit to the northern
hemisphere data alone.  The dashed grey line is the power in the
Bianchi template alone.\label{fig:cls_compare}}
\end{centering}
\end{figure}

We have verified, however, that changing the assumed parameters and
using, for example, a flat $Q=18\,\mu$K power law spectrum, or a completely
implausible spectrum, has little effect (less than $3\%$) on the
resulting best-fit amplitude and position for the Bianchi
component. In fact, the power spectrum affects only the estimated
significance of the result, as that significance is dependent on the
expected level of large-scale CMB structure that drives chance
alignments.  As shown in Figure \ref{fig:cls_compare}, correction for
this Bianchi model lowers the large-scale power.  Our significance
estimates are based on simulations generated assuming a higher level
of large scale power, so the significance of the detection would
increase when compared to an ensemble consistent with the corrected
power spectrum.

\section{Implications}

There are several interesting results based on \emph{WMAP} first-year
data that are inconsistent with the assumptions of isotropy and
Gaussianity and that are immediately relevant to this study.  De
Oliveira-Costa \etal (2004), \citet{land:2005}, and \citet{copi:2005}
(and sources therein) examine the low-$\ell$ multipoles of the
foreground-cleaned data and find that, in addition to the anomalously
low quadrupole amplitude, the preferred axes of the quadrupole and
octopole are anomalously well aligned in the direction of
$(l,b)=(-110\degr,60\degr)$.  
\citet{eriksen:2004a} and \citet{hansen:2004a} find a system of
reference (roughly aligned with the ecliptic) in which there is a
significant difference in large-scale power between the two
hemispheres at the 98\%-99\% level, with significantly more power in the
south.
\citet{vielva:2004} and \citet{cruz:2005} detect non-Gaussianity in
the \emph{WMAP} combined Q-V-W map using spherical wavelets; they find
significant kurtosis in the wavelet coefficients at a scale of
$10\degr$ and identify a cold spot at $(l,b)=(209\degr,-57\degr)$ as
the probable source.  Our choice of models was partly motivated by the
morphology of these anomalies, and indeed, subtracting for our
best-fit Bianchi template corrects them.

\subsection{Quadrupole Amplitude}\label{sec:quadrupole}

The quadrupole amplitude has been considered anomalously low since
\emph{COBE} (see \citealt{de Oliveira-Costa:2004} and references
therein).  As pointed out by \citet{jaffe:2005}, the correction for
this Bianchi component raises the low quadrupole amplitude to a value
more consistent with the theoretical power spectrum.  This result is
unchanged with the best-fit model of this work, since the models are
almost the same.  Should this be considered ``fine tuning''?  

We can simulate the situation by taking as the the primordial
quadrupole the \emph{WMAP} quadrupole (as derived by
\citealt{bielewicz:2004}) minus the quadrupole of our best-fit Bianchi
model.  If we then add the Bianchi quadrupole at random orientations,
we can see how likely it is that the resulting total quadrupole be as
low as the observed \emph{WMAP} quadrupole.  We find that the
likelihood is
$\sim5\%$.  This implies that the level of ``fine tuning'' required to
end up with the low observed quadrupole is not exceptional.

We further take a set of 1000 simulated Gaussian CMB skies, with and
without a Bianchi component, and fit our best-fit Bianchi model to
them.  The ``corrected'' quadrupole is on average $\sim5\%$ lower than
the original, which is to be expected considering that the fit is a
least squares solution. In contrast, using the real LILC data, the
correction has the effect of raising the quadrupole.  This happens in
over $\sim20\%$ of the simulations, so while this is not the average
behavior, it is not extraordinary.

%

\subsection{Low-$\ell$ Alignment and Planarity}\label{sec:lowl}

\begin{figure}
\plotone{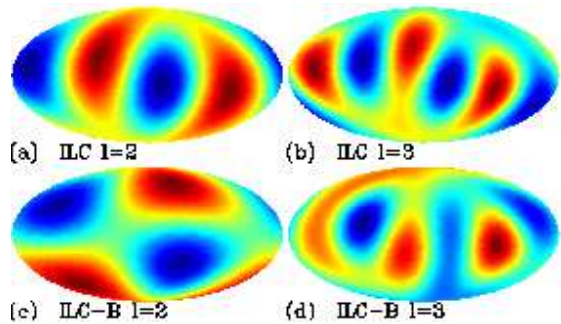}
\figcaption{\small Low-$\ell$ multipoles of the WILC corrected ({\it bottom panels}) and uncorrected ({\it top panels}) for the Bianchi component. \label{fig:low_ls}}
\end{figure}

De Oliveira-Costa et al.\ (2004), \citet{land:2005}, and
\citet{copi:2005} discuss the statistically anomalous alignment of the
quadrupole and octopole in the \emph{WMAP} data.  The preferred axes
of the $\ell =2$ and $\ell =3$ multipoles are only $7\degr$ apart
(roughly in the direction of $(l,b)=(-110\degr,60\degr)$), which is
anomalous at the $99.3\%$ level compared to simulations.  After
subtracting the best-fit Bianchi template, these axes lie $74\degr$
apart, consistent (at $27\%$) with the statistically isotropic
simulations (see Fig. \ref{fig:low_ls}).

The planarity of the low-$\ell$ multipoles has also been considered
somewhat anomalous (see \citealt{de Oliveira-Costa:2004} and
\citealt{land:2005} for a discussion).  The {\it t}-statistic defined by
\citet{de Oliveira-Costa:2004} provides a measure of this planarity.
Again, subtracting the Bianchi template lowers the significance of the
the low-$\ell$ multipoles.  The planarity of the octopole in
particular drops from a significance of $\sim 90\%$ (depending on
whether the WILC or LILC is used) to $\sim 50\%$.  Figures
\ref{fig:low_ls} (b) and (d) shows how the planarity of the octopole
is disrupted.  This will also impact the results of multipole vector
analyses such as that of \cite{copi:2005}.

\subsection{Large-Scale Power Asymmetry}\label{sec:asymmetry}

\begin{figure} \centering
\plotone{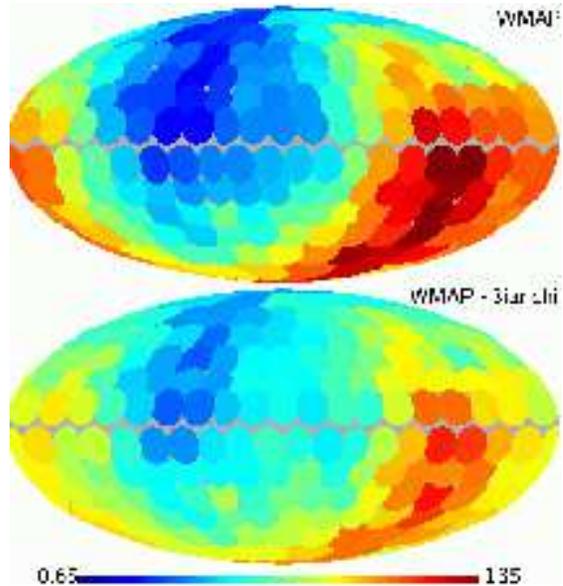}
\figcaption{\small Power ratio between hemispheres in \emph{WMAP} ILC, corrected (\emph{bottom}) and uncorrected (\emph{bottom}) for the best-fit Bianchi component.
\label{fig:power_asymmetry}}
\end{figure}

\citet{eriksen:2004a} and \citet{hansen:2004a} reported that the
large-scale power ($\ell \lesssim 40$) in the \emph{WMAP} data is
anisotropically distributed over two opposing hemispheres (in the
reference frame in which the z-axis points toward
%
$(l,b)=(57\degr,10\degr)$; see Fig. \ref{fig:power_asymmetry}), with
a significance of $3\sigma$ compared with simulations.  Repeating the
analysis and adopting the Kp2 sky coverage, we compare the corrected
V+W \emph{WMAP} map with 2048 simulations. We find that $\sim 14\%$ of
the simulations have a larger maximum power asymmetry ratio than the
Bianchi-corrected map, whereas only $0.7\%$ have a larger ratio than
the uncorrected data (see Fig.
\ref{fig:power_asymmetry}).  It is apparent that the maximum power
ratio between any two hemispheres is significantly suppressed after
subtracting the Bianchi template, as no asymmetry axis is found at any
statistically significant level.  It is apparent from that figure,
however, that some residual power asymmetry remains.  This comes
largely from the range $20<l<40$, where the Bianchi template has little
power, indicating that a model with more small-scale structure may be
needed.

\subsection{Wavelet Kurtosis}\label{sec:wavelet}

\citet{vielva:2004} and \citet{cruz:2005} used a wavelet technique to detect an excess of
kurtosis in the wavelet coefficients and isolate an unusually cold
spot ($\sim 3\sigma$ significance relative to Gaussian simulations) at
Galactic coordinates $(l,b)=(209^{\circ},-57^{\circ})$.  Referring
again to Figure
\ref{fig:ilc_b_bilc}, we see that a cold spot is indeed present at the
right location, in the form of the center of the spiral.

We therefore also repeat the analysis of \citet{vielva:2004}, and
compute the kurtosis of the wavelet coefficients as a function of
scale from both the WILC and the corresponding Bianchi-subtracted
map. A $|b|<20\degr$ galactic cut is imposed in this case, for
computational convenience.

\begin{figure}
\plotone{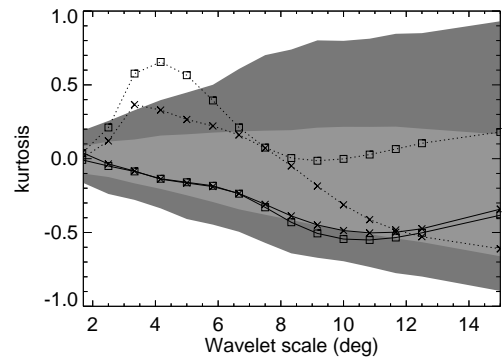}
\figcaption{\small Kurtosis in wavelet coefficients. The  boxes and
 crosses show the kurtosis before and after subtracting the
Bianchi template, respectively, computed from the southern ({\it
dotted line}) and northern ({\it solid line}) Galactic hemispheres.\label{fig:kurtosis}}
\end{figure}

The results from this exercise are reported in Figure
\ref{fig:kurtosis} After subtracting the Bianchi template, the significance
of the southern hemisphere anomaly is greatly reduced, and no new
non-Gaussian features have been introduced.

%

\section{Discussion and Conclusions}

We have considered a fast and efficient method for fitting a template
to the full sky in harmonic space and finding the best-fit location
and orientation.  The total convolver algorithm evaluates the
correlation between the sky and the template at every possible
relative orientation using fast Fourier transforms.  With this
algorithm, the search for the best-fit becomes 2 orders of magnitude
faster than the corresponding search performed one rotation at a time.
This method, along with pixel-space simultaneous foreground fitting,
provides a powerful tool for testing any deterministic model for
anisotropy in the CMB.  Simulations generated by the LILC pipeline
allow us to quantify the bias, investigate the effects of
foreground contaminants, and show how well each of these methods
detects a known input.

We have applied this method to the first-year \emph{WMAP} data to
search for evidence of shear and vorticity using templates derived for
Bianchi type VII$_{h}$ universes.  We find a surprisingly significant
correlation between the \emph{WMAP} data and a right-handed Bianchi model
with $x=0.62$, $\Omega_0=0.5$, and shear of
$\left(\frac{\sigma}{H}\right)_0=4.3\pm 0.8\times 10^{-10}$,
implying a vorticity of
$\left(\frac{\omega}{H}\right)_0=9.5\times10^{-10}$.  The center of
the spiral structure lies at approximately
$(l,b)=(222\degr,-62\degr)$.  Simulations show that this amplitude is
likely to be biased by $\sim7\%$, implying a true amplitude closer to
$4.0\times10^{-10}$.  Incomplete sky fits, simultaneous foreground
fitting, and fits to a set of Gibbs samples are all consistent with
this amplitude and indicate that confusion with Galactic emission is
unlikely to contribute significantly to this detection.

Correcting the \emph{WMAP} data for the effect of the best-fit model
solves several problems seen in the data.  The corrected maps show
significantly reduced power asymmetry between any two hemispheres.
The correction also eliminates the non-Gaussian kurtosis in the
wavelet coefficients detected by
\citet{vielva:2004} and \citet{cruz:2005}, raises the low measured
quadrupole by a factor of 2, and disrupts the planarity of the
octopole and its anomalous alignment with the quadrupole.  In short,
the data appear far more Gaussian and isotropic after correction.

The original analyses by \citet{kogut:1997} and \citet{bunn:1996} were
limited by the signal-to-noise ratio level in the DMR instrument.  Our best-fit
 result is just under their upper limit but still significant due
to \emph{WMAP}'s greatly improved signal-to-noise ratio.  Furthermore, the
Kogut analysis searched a coarse ($\sim 10\degr$) grid of possible
locations and orientations, while with the total convolver, we can
efficiently search a finer grid.  

How likely is it that our best-fit model is a true detection rather
than a chance alignment?  Considering the best-fit model by itself and
comparing its fit amplitude to simulations, it is higher than $99.7\%$
of simulations.  However, the simulations also show that $10\%-20\%$ of
Gaussian, statistically isotropic skies will have one of the Bianchi
models appear as significant.  Considering the fact that the sky is
approximately Gaussian and isotropic, one would not expect to find a
more definitive detection based on template fitting alone.  But the
distribution of chance alignments in the simulations is sensitive to
the amount of large-scale power assumed, and that is lowered by the
Bianchi correction to the \emph{WMAP} data.  Furthermore, the
cumulative probability that a chance alignment not only fits at the
level of our best-fit model but also has the effect of resolving the
several anomalies in the data must also be considered in any
qualitative judgment of the significance of this result.

Further improvement to the data will not refine these measures
significantly, because at the \emph{WMAP} sensitivity level, the
analysis is already very close to the expected distribution of chance
alignments in the absence of noise.  Improved foreground subtraction
will, however, remove some of the possible confusion and bias,  but
neither higher resolution nor higher signal-to-noise ratio data should
change this result nor be able to provide additional information
concerning the question of whether the fit is a real detection of
vorticity and shear.  Answering that question will require additional
verifiable predictions for the effects of vorticity and shear on other
observables.

However, in the context of the anomalies that this hypothesis can explain,
the possible detection is certainly provocative.  The most important
result of this analysis is that a model with vorticity and shear can
explain the observed asymmetry in the CMB anisotropies and the
non-Gaussian cold spot.  Note that this asymmetry exists only in the
$\Omega_0<1$ versions of these Bianchi models.  Significant evidence
currently indicates that $\Omega_0$ is very close to 1, so our best-fit
 model cannot be considered physically realistic.  However, as
mentioned in \S\ \ref{models}, Barrow \etal (1985) did not include any dark
energy component.  Furthermore, the Bianchi model does not include a
mechanism to generate structure at the surface of last scattering.  A
self-consistent theory is required that can explain the small scale
fluctuations, and in particular the acoustic peaks, in the context of
a Universe with shear and vorticity.  But from a pragmatic point of
view, one can conclude that, regardless of the viability of the
particular Bianchi model, this result gives a measure of the
significant deviation from isotropy in the data.

We consider this result to be further motivation for considering ideas
outside of the so-called concordance model of cosmology.  There are
anomalies in the data that are inconsistent with the theory of a
Gaussian, statistically isotropic universe, and Bianchi models are
only one such anisotropic model that merits investigation.  We have
demonstrated a method of template fitting that can be applied to test
any model that makes a deterministic prediction for an anisotropy
pattern in the CMB.  The best-fit Bianchi model provides a template
temperature pattern that can explain the observed anomalies in the
data and that describes the morphology theorists may need to reproduce
in considering alternatives to the standard cosmological model.

\section*{Acknowledgments}

We are grateful to M.~Demianski, S.~Hervik, and S.~D.~M.~White for
useful discussions, and to J.~McEwen for finding errors in the
original numbers.  H.~K.~E. acknowledges financial support from the
Research Council of Norway, including a Ph.\ D. scholarship. F.~K.~H. was
supported by a Marie Curie Re-integration Grant within the 6th
European Community Framework Programme.  We acknowledge use of the
HEALPix software 
\citep{healpix} and analysis package for deriving the results in this
paper.  We also acknowledge use of the Legacy Archive for Microwave
Background Data Analysis (LAMBDA).

\end{document}